# Far-field nanoscale infrared spectroscopy of vibrational fingerprints of molecules with graphene plasmons

Hai Hu[1*], Xiaoxia Yang[1*], Feng Zhai[2], Debo Hu[1], Ruina Liu[1], Kaihui Liu[3], Zhipei Sun[4], Qing Dai[1]

[1] National Center for Nanoscience and technology, Beijing 100190, China,

[2] Department of Physics, Zhejiang Normal University, Jinhua 321004, China,

[3] School of Physics, Center for Nanochemistry, Collaborative Innovation Center of Quantum Matter, Peking University, Beijing 100871, China,

[4] Department of Micro- and Nanosciences, Aalto University, Tietotie 3, FI-02150 Espoo, Finland

*These authors contributed equally to this work.

Correspondence: daiq@nanoctr.cn; zhipei.sun@aalto.fi; khliu@pku.edu.cn.

## Abstract:

Infrared spectroscopy, especially for molecular vibrations in the fingerprint region between 600 and 1500 cm$^{-1}$, is a powerful characterization method for bulk materials. However, molecular fingerprinting at the nanoscale level still remains a significant challenge, due to weak light-matter interaction between micron-wavelengthed infrared light and nano-sized molecules. Here, we demonstrate molecular fingerprinting at the nanoscale level using our specially designed graphene plasmonic structure on CaF$_2$ nanofilm. This structure not only avoids the plasmon-phonon hybridization, but also provides *in situ* electrically-tunable graphene plasmon covering the entire molecular fingerprint region, which was previously unattainable. In addition, undisturbed and highly-confined graphene plasmon offers simultaneous detection of in-plane and out-of-plane vibrational modes with ultrahigh detection sensitivity down to the sub-monolayer level, significantly pushing the current detection limit of far-field mid-infrared spectroscopies. Our results provide a platform, fulfilling the long-awaited expectation of high sensitivity and selectivity far-field fingerprint detection of nano-scale molecules for numerous applications.

## Introduction

Far-field mid-infrared (MIR) spectroscopy plays an increasingly important role for numerous applications (for example, chemical detection[1], food safety[2], and bio-sensing[3]) through directly probing vibrational characteristics of a broad range of molecular species and compounds. In particular, in the molecular fingerprint region[4-6] from 600 to 1500 cm$^{-1}$ (which corresponds to the wavelength range of approximately 6-16 μm)[4], complex vibrational characteristics of molecules in bulk materials can be effectively distinguished to enable unambiguous identification of molecular structures and species[4-6]. Currently, high sensitivity and selectivity molecular detection at the nanoscale level in the fingerprint region is in great demand for various applications. However, as the wavelengths in this spectral range (approximately 10 μm) approach lengths more than three orders of magnitude larger than typical molecules (smaller than 10 nm), the light-molecule interaction cross-section becomes remarkably small[7-9]. Further, unique but complicated patterns of vibrational fingerprint modes produced by each different molecular components make selective identification difficult. These therefore yield extremely low sensitivity and selectivity detection of molecular vibrational fingerprints using infrared (IR) spectroscopy, which is insufficient to address ongoing challenges and emerging applications (for example, trace chemical detection and sensing for safety and health applications).

Plasmonic enhancement, one of the most effective approaches to improve light-matter interaction, has been implemented to enhance light-molecule interaction in the MIR range[4,10-27]. However, in contrast to its successful implementation in the visible and near-infrared spectral range, noble metal plasmons cannot fully fulfill the enhancement requirement in the molecular fingerprint region due to its weak field confinement, narrow spectral resonance and very limited tunability[4,11-13]. Recently, graphene has been demonstrated to manipulate electromagnetic signals at deep-subwavelength scale with ultrahigh field confinement due to its unique electronic band structures[28-34]. Especially, graphene plasmon resonance can be tuned by the structure dimensions and electrostatic doping in the terahertz and IR region[35-39]. These extraordinary properties make graphene plasmon a promising candidate for detection enhancement applications in the MIR region, which has been identified in recent experimental achievements[6,8,40-45]. However, in the technologically important MIR fingerprint region, graphene plasmon applications are severely restrained by strong coupling between the graphene plasmons and substrate phonons[46-54], which strongly confines the electromagnetic energy between graphene and substrate. This results in low near-field enhancement on the graphene top surface and a limited spectral tunability, which are not suitable for sensing applications (for example, molecule fingerprint sensing).

Here, we demonstrate a hybrid graphene plasmonic structure on $CaF_2$ nanofilm (graphene/$CaF_2$), in which there is no substrate phonon effect in the MIR region. This eliminates the strong plasmon-phonon coupling issue in conventional graphene plasmonic structures, and enables the first electrically tunable plasmon that covers the entire molecular fingerprint region for molecular detection with extremely high sensitivity down to sub-monolayer level. Highly-selective average enhancement of up to larger than 20-fold was achieved for unambiguous identification of vibrational fingerprints in polymers at the nanoscale with a far-field Fourier transform infrared spectroscopy. In addition, our graphene plasmonic devices can simultaneously identify in-plane and out-of-plane vibrational modes in complicated molecules. Our results provide a route for highly sensitive and selective identification of molecular structure fingerprints for diverse applications in anti-terrorism, food and healthcare, and bio-sensing.

## Results

### Graphene plasmonic device in the fingerprint region

The design of our graphene-based molecular fingerprint sensor is illustrated in Fig. 1a. We introduce a 300-nm $CaF_2$ thin-film based dielectric supporting substrate for graphene (Figs 1b and 1c), due to its broadband transparency in the IR spectral region and suitable dielectric properties[55]. As shown in Fig. 2a (also Supplementary Figure 1), $CaF_2$ does not exhibit any active phonon mode in the IR region and provides a constant background in the fingerprint region of 675-1500 $cm^{-1}$, which is ideal for *in situ* spectroscopy in the MIR region[10,38]. In comparison, conventional $SiO_2$ and hexagonal boron nitride (h-BN) substrates produce strong phonon-related resonance peaks in this spectral region (Fig. 2a). These phonons can strongly hybridize with graphene plasmon, and drastically reduce graphene plasmonic field enhancement and electrostatic tunability[47,48]. Further, in contrast to bulk $CaF_2$ substrates[10,38] and diamond-like carbon substrates[48], our $CaF_2$ nanofilm is an excellent dielectric material for highly efficient graphene electrical doping[56]. Electrically continuous graphene nanoribbon arrays (Fig. 1b) connected with Au electrodes were fabricated on the $CaF_2$/Si substrate (see Methods and Supplementary Figures 2-5 for further details). The electrostatic tunability of our graphene/$CaF_2$ device is shown in Fig. 1c. The graphene carrier density can be tuned up to $2.6 \times 10^{12}$ cm$^{-2}$ with a relatively low gate voltage ($V_g$= -15 V, as plotted in Fig. 1c), which demonstrates effective dielectric isolation of the $CaF_2$ thin film. The calculation of the carrier density and Fermi level is detailed in Supplementary Note 1.

A far-field Fourier transform infrared microscope (FTIR) was used to characterize the plasmonic response of graphene nanoribbons (see Methods for further details). Figure 2b shows the extinction spectra of a graphene nanoribbon array with an 80 nm ribbon width at various Fermi levels. As shown, the extinction spectra of our graphene/$CaF_2$ device exhibit a single but prominent peak, while the graphene/$SiO_2$ devices show multiple peaks due to the hybridization of plasmon and phonons (Supplementary Figures 6-8). This indicates that the excellent intrinsic properties of the graphene plasmons were preserved in our graphene/$CaF_2$ device without disturbance by the substrate phonons. By changing the graphene Fermi level within the gate voltage change of approximately 21 V, the plasmon resonance peak can be tuned over the 900-1400 cm$^{-1}$ region. We can cover shorter and longer wavelength regions using graphene devices with different ribbon widths, as shown in Supplementary Figures 6 and 7. The graphene plasmon field confinement factor (defined by the wavelength ratio between the excitation light and graphene plasmon) was calculated at greater than 50, which indicates a large field enhancement in our graphene plasmonic molecular fingerprint sensors.

**High sensitivity and selectivity molecular fingerprinting**

To demonstrate the excellent performance of our molecular fingerprint sensors, we used ultrathin (8-nm-thick) polyethylene oxide (PEO) films because IR spectra of PEO bulk films are well-studied and feature fourteen well-characterized molecular vibrational modes in the fingerprint region[57]. As shown in the pristine absorption spectrum (black curve in Fig. 3a) of an ultrathin PEO film (that is, without the graphene plasmon enhancement), the molecular vibrational absorption is weak (less than 0.02% nm$^{-1}$). Most absorption peaks cannot be distinguished from the background noise, which is expected from weak light-molecule interaction in this spectral range. In striking contrast to the noisy results without plasmon enhancement, the extinction spectrum (red curve in Fig. 3a) of the same PEO film covering the graphene nanoribbon array clearly features various strong dips over the broad graphene plasmon resonance after enhancement. These dips were assigned to fourteen molecular fingerprints of the PEO molecules, which correspond to various rocking, stretching, wagging, and twisting molecular vibrational modes (Fig. 3b). These modes correlate well with previous IR absorption results on bulk PEO films (Supplementary Figure 8). The dip behaviors originate from destructive interference when the graphene plasmons and vibrational modes interact with an opposite phase relationship. We also compared the sensing results obtained with a SiO$_2$ substrate (Supplementary Figure 9), which confirms that our CaF$_2$ nanofilm indeed provides significant signal enhancement, due to the elimination of the strong plasmon-phonon coupling issue in conventional graphene plasmonic structures.

The vibrational mode signals close to the graphene plasmon resonance peak feature a much larger amplitude than these wing signals. For instance, the strength (defined by the peak area) of the mode F located at 1123 cm$^{-1}$ (the strongest peak in the pristine polymer spectrum, which corresponds to the anti-symmetric stretching mode in the C-O-C group of the PEO molecules) has been enhanced by more than 10 times (calculation details in Supplementary Figure 10 and Supplementary Note 2). Considering that this is the average enhancement for the PEO film over the all detection area (0.2×0.1 mm$^2$) with a 1:3 graphene ribbon-to-pitch ratio, and the plasmonic field enhancement is strongly concentrated at the ribbon edges, the enhancement factor at the ribbon edges should be substantially larger than the average enhancement factor. We also tested the reusability of our plasmonic sensors (Supplementary Figure 11 and Supplementary Note 3), which shows that our graphene devices are reusable and robust.

The graphene plasmon broadband electrical tunability offers unique frequency-selective enhancement for individual vibrational modes in our molecular fingerprint sensor as demonstrated by the plasmon-enhanced molecular fingerprinting results (Fig. 4a). For example, the graphene plasmon resonance frequency can be continuously tuned from 1052 to 1198 cm$^{-1}$ by changing $\Delta_{CNP}$ from -6 to -21 V. The signal strengths of vibrational modes A and B gradually decrease when $\Delta_{CNP}$ decreases from -6 to -21 V, which is expected because the graphene plasmon resonance frequency is tuned away from these modes. However, because the graphene plasmon resonance frequency is tuned towards the H, I, J and K modes, the signal intensities of these modes rapidly increase with a decrease in $\Delta_{CNP}$. To further investigate this frequency-selective function quantitatively, we fit the fourteen vibrational modes, extracted their peak areas, and calculated their enhancement factors based on our experimental results (calculation details in Supplementary Note 2)[58]. The calculated enhancement factors are shown in Fig. 4b (and Supplementary Figure 12) as a function of the frequency difference ($\Delta f$) between the vibrational mode and graphene plasmon. As expected, these fitted curves follow the fundamental principle of plasmonic enhancement: molecular signal enhancement ($\Delta I$) is inversely proportional to the frequency detuning $\Delta f$ ($\Delta I \sim 1/\Delta f$)[40]. This result demonstrates that the electrically tunable graphene plasmon can selectively enhance the vibrational modes. For instance, the overall enhancement factor of the mode G (at 1151 cm$^{-1}$, an interaction mode between the symmetric stretching of the C-C group and anti-symmetric stretching of the C-O-C group in the PEO molecules) can be adjusted from 2.5 to 23 via graphene plasmon frequency tuning.

In addition, with the electrical tunability enhancement, we can distinguish the overlapping vibrational modes in the pristine absorption spectrum. For example, the E and F modes cannot be distinguished in the absorption spectrum (Fig. 4c). However, the enhancement of modes E and F can be well modulated by electrically adjusting the graphene plasmon resonance frequency. In contrast to noble metal surface plasmons, which are typically tuned by changing the dimensions[58], graphene plasmons with electrical tunability are superior for *in situ* frequency-selective enhancement and thus facilitate unambiguous identification of complex molecular compounds and structures through correlation with structural vibrational fingerprint detection.

## Simultaneous detection of in-plane and out-of-plane vibrational modes

Here, we demonstrated that our molecular fingerprint sensor can simultaneously perceive in-plane and out-of-plane vibrational modes. The out-of-plane modes (also named as ZO modes) are not typically visible for conventional IR microscopy because these vibrational modes cannot respond to incident light with a polarization direction perpendicular to its vibrational direction[57]. To clarify this advantage, we choose h-BN monolayer as an example, which has both in-plane and out-of-plane IR modes[59]. Figure 5a shows the plasmonic enhanced IR spectra of h-BN monolayer as well as its pristine IR absorption spectrum. The latter one only features the in-plane structure vibrational mode at approximately 1370 $cm^{-1}$ and the out-of-plane structure vibrational mode at approximately 820 $cm^{-1}$ cannot be detected using the conventional IR microscopy[59] (Supplementary Figure 13). However, with the graphene plasmon enhancement, both the out-of-plane and in-plane vibrational modes in h-BN can be clearly detected (Fig. 5a). The sensing sensitivity (that is, the dip amplitude) is greatly enhanced (purple curve for a larger out-of-plane vibrational mode enhancement) with a good match between the graphene plasmonic resonance and vibrational mode after tuning the electrical gate. Such superior performance in detecting both the in-plane and out-of-plane vibrational modes is explained by the graphene plasmon enhancement effect in these directions[51], as depicted schematically in Fig. 5b. As shown, the electric fields of graphene plasmons can follow along both the $x$ and $z$ directions, and their enhancements are comparable (Supplementary Figure 13). Therefore, the in-plane and out-of-plane molecular vibrational modes can be possibly detected in our plasmonic sensor. Theoretical explanation on the coupling mechanism between the out-of-plane mode and the graphene plasmon is also given with a classic electromagnetic coupling theory in Supplementary Note 4. This unique property, wherein multi-directional vibrational modes are accessed (that is, the out-of-plane and in-plane modes) provides a promising approach for probing sophisticated changes in molecular configurations, such as protein folding or molecular crystallization at interfaces.

**Detection of sub-monolayer polymer**

In principle, highly confined graphene plasmons can facilitate high-sensitivity, sub-monolayer detection. Based on near-field distribution, we calculated near-field intensity confinement as a function of a given distance $d$ from the graphene ribbons (Fig. 6a). The data show that approximately 65% of the plasmon energy is confined within a 0.5 nm gap from the graphene surface, which suggests possible plasmon enhancement for sensing trace elements with dimensions as small as 0.5 nm. For verification, we performed sensing experiments on a residual polymethyl methacrylate (PMMA) polymer during the device fabrication process of our graphene device. Generally, a PMMA monolayer is approximately 4 nm thick[60], but a residual PMMA layer is typically a discontinuous sub-monolayer film with a maximum thickness less than 1.5 nm (Supplementary Figure 3). Such sub-monolayer films cannot be detected through conventional IR microscopy due to its low sensitivity. The extinction spectra of the residual PMMA on our molecular fingerprint sensor at varied effective gate voltages are shown in Fig. 6b. As shown, the PMMA vibrational fingerprint modes at 1151, 1192, 1244, and 1269 cm$^{-1}$, can be identified in the prominent plasmon resonant peaks[61]. Because the amount of the PMMA molecules is extremely small, their IR signals become very weak and even invisible when the detuning between molecular vibrational peaks and the plasmon peak is large (for example, when $\Delta_{CNP}$ = -3 V). Thanks to the electrical tunability of graphene plasmon, we can tune its resonance with the gate voltage to selectively identify the vibrational fingerprints of sub-monolayer polymer (for example, the ones corresponding to $\Delta_{CNP}$ of -9 V and -15 V). These observations clearly indicate that our molecular fingerprint sensor is highly sensitive and selective even for sub-monolayer molecules detection.

## Discussion

In conclusion, we demonstrate high sensitivity and selectivity molecular fingerprinting with a graphene plasmon based far-field Fourier transform infrared spectroscopy. We enhance approximately 8-nm PEO film vibrational signatures up to larger than 20-fold. We also show that our method can detect structural vibrational modes (both the in-plane and out-of-plane structural vibrations) in molecules with a resolution previously unachievable using conventional mid-IR absorption spectroscopy. Our graphene plasmon based far-field molecular fingerprint spectroscopy can facilitate various applications, such as functional studies on trace matter in biology and pharmacology.

## Methods

### Device preparation

Graphene and monolayer h-BN were grown on copper foil through chemical vapor deposition (CVD) methods and transferred onto a SiO$_2$ (285 nm)/Si substrate using the PMMA-assisted method. Nanoribbon arrays were patterned onto the graphene surface using electron beam lithography (EBL) (Vistec 5000+ES, Germany) in PMMA followed by oxygen plasma etching. An additional EBL process and electronic beam evaporation were used to define the electrodes. We deposited a 300-nm thick CaF$_2$ film onto a low-doped Si substrate using electronic beam evaporation. The complete graphene device was then transferred onto the CaF$_2$/Si substrate. A thin (approximately 8 nm) PEO polymer layer was spin-coated onto the sensor as a sensing sample.

**Characterization**

Patterned graphene nanoribbons were characterized with scanning electron microscopy (SEM, Hitachi S-4800) and atomic force microscopy (AFM, Neaspec s-SNOM). The graphene quality was confirmed by Raman spectroscopy (Horiba Jobin Yvon LabRAM HR800) before and after oxygen ion bombardment under PMMA protection. The electrical transport properties were characterized using a semiconductor parameter analyzer (Agilent 4294A) at room temperature.

**FTIR microscopy measurements**

IR transmission measurements were performed using a FTIR microscopy (Thermo Fisher Nicolet iN10). We generated a background spectrum for each measurement. A bare CaF$_2$/Si substrate was used to extract the background signal from the pristine PEO film absorption spectrum. The graphene nanoribbon array transmission spectra at CNP ($T_{CNP}$) were used as background, and the transmission spectra ($T_{EF}$) at certain Femi levels $E_F$ were collected to obtain the plasmonic extinction spectra by 1-$T_{EF}/T_{CNP}$. Thus, the extinction spectra are purely plasmonic signals excited by the incident light. Each measurement was repeated several times to confirm the extinction spectrum.

**Simulation**

The graphene plasmons were simulated by using the finite element method. The IR lights impinge perpendicular on graphene nanoribbons. We modeled graphene as a thin film with a thickness of $t$ and imposed the relative permittivity $\varepsilon_G = -i\sigma/\omega\varepsilon_0 t$. $\sigma(\omega)$ is the complex optical conductivity of graphene evaluated within the local random phase approximation[62]. The graphene thickness is set to be 1 nm, at which the calculations reach proper convergence.

**Data availability**

The data that support the findings of this study are available from the corresponding author upon request.

## Acknowledgements

We thank Javier Aizpurua and Tomáš Neuman for useful discussion of the theoretical simulation. Z.S. thanks Amos Martinez for his comments on the manuscript. This work was supported by the National Basic Research Program of China (Grant No. 2015CB932400), the National Natural Science Foundation of China (Grant No. 51372045, 11504063, 11174252, 11474006, 51522201, and 91433102), the Academy of Finland (Grant No: 276376, 284548, 295777), TEKES (OPEC), and the European Union's Seventh Framework Programme (REA grant agreement No. 631610).

## Author contributions

Q.D. and X.Y. conceived the experiments; H.H. performed the device fabrication, characterization, and FTIR measurements; F.Z. provided modelling and the theoretical studies; D. H. performed AFM characterization and Near-field IR image measurements; R.L. helped in device fabrication and FTIR measurements; H.H. X.Y., Q.D., Z.S., and K.L analyzed the data and co-wrote the manuscript.

## Competing financial interests

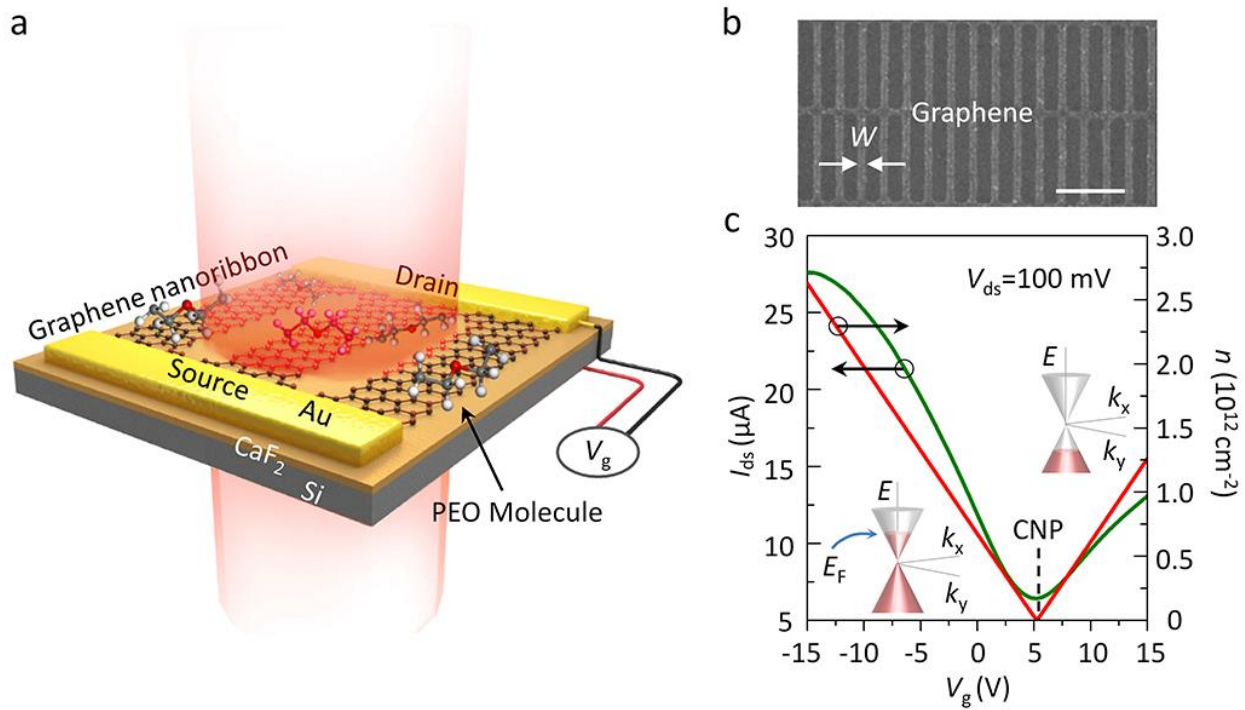

**Figure 1. Graphene plasmon enhanced molecular fingerprint sensor. (a)** A schematic of the sensor. The graphene nanoribbon structure was designed on a $CaF_2$ dielectric substrate (300 nm thick). The graphene plasmon resonance excited by the incident IR beam (the red shaded pillar) can be tuned *in situ* by electrostatic doping through the gate voltage ($V_g$). **(b)** A scanning electron microscope (SEM) image of the graphene nanoribbon pattern. Ribbon width ($W$): 80 nm; width-to-pitch ratio: 1:3. Scale bar, 1 μm. **(c)** The transfer curve (green line) of our graphene/$CaF_2$ fingerprint sensor. The gate voltage that corresponds to the charge neutral point (CNP, $V_{CNP}$, marked as dash line) is approximately 5 V.

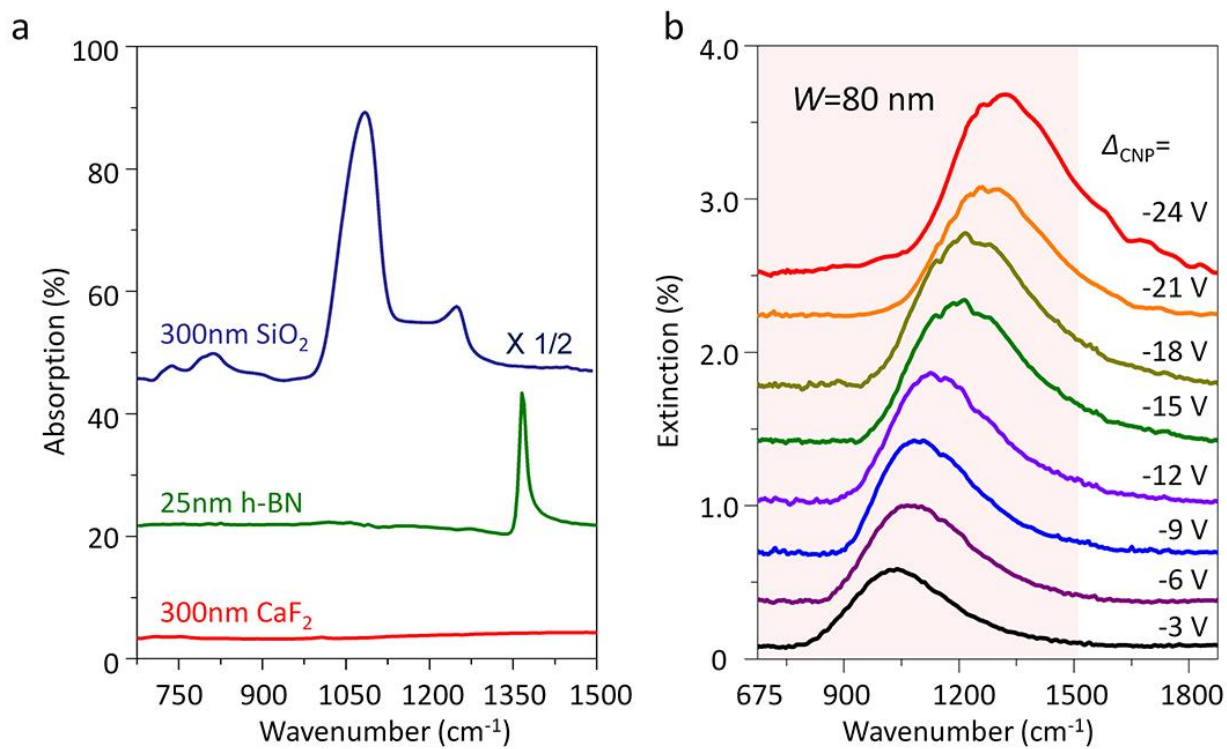

**Figure 2. Optical properties of CaF₂ nanofilm substrate. (a)** A comparison of the IR absorption spectra of various dielectric substrates ($SiO_2$, h-BN, and $CaF_2$). **(b)** The broadband tunable intrinsic graphene plasmon in the graphene/$CaF_2$ fingerprint sensor via the effective gate voltage ($\Delta_{CNP} = V_g - V_{CNP}$). The red shaded region indicates the molecular fingerprint region.

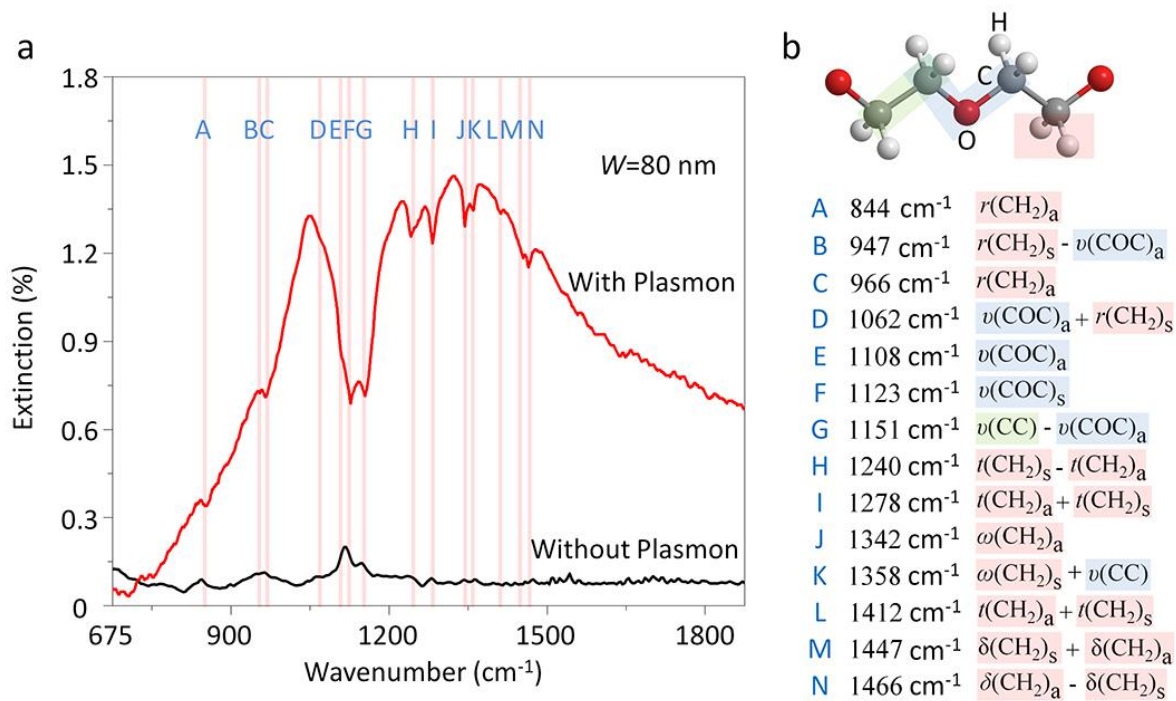

**Figure 3. Highly sensitive detection of molecular vibrational fingerprints. (a)** A comparison of the sensing results for an 8-nm-thick PEO film with (red curve) and without (black curve) graphene plasmon enhancement. The corresponding Fermi level is approximately 0.2 eV. The red vertical lines indicate various PEO molecular vibrational modes. **(b)** The list of PEO vibrational modes in the molecular fingerprint region and their positions in **(a)**. The green, blue and red backgrounds represent the C-C, C-O-C, and methylene groups, respectively. The prefixes **r**, **υ**, **ω** and **t** indicate rocking, stretching, wagging and twisting modes, respectively. The suffixes **s** and **a** imply symmetric and anti-symmetric modes, respectively, with respect to the two-fold axis perpendicular to the helix axis and passing through the oxygen atom or center of the C-C bond. The + and - signs denote the phase relationship for the potential energy distribution of the coupled coordinates.

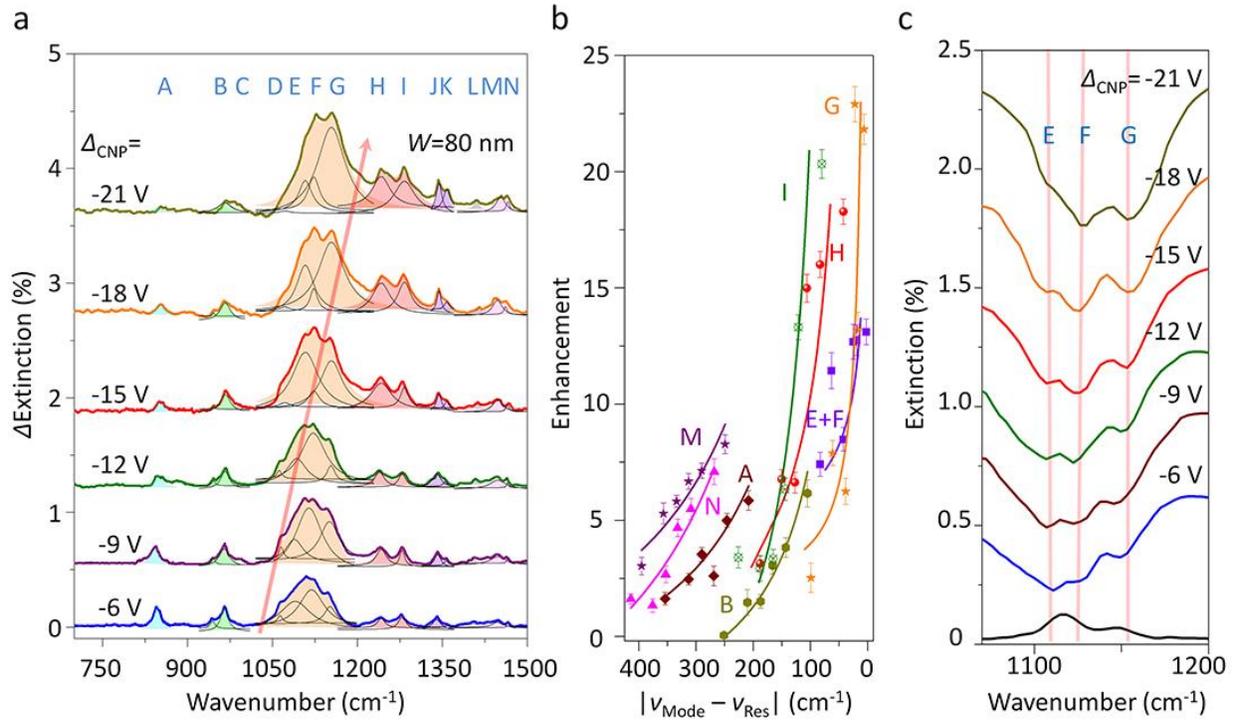

**Figure 4. Highly selective detection of molecular vibrational fingerprints. (a)** The plasmon-induced vibrational mode response of the PEO molecules extracted from the extinction spectra of the graphene plasmonic resonance peak at different effective gate voltages. The change of the center frequency of each plasmon peaks is indicated by the red arrow. The Lorentz line shapes (the thin solid curves) are used to fit the peaks induced by different vibrational modes of PEO molecules. The shaded areas with different colors indicate the superposition of the fitted peak areas. **(b)** The enhancement factor of typical vibrational modes as a function of the distance between the mode ($\nu_{Mode}$) and graphene plasmon resonance peak ($\nu_{Res}$). The error bars in the plots are standard deviation from large numbers of measurements. **(c)** Enlarged extinction spectra for the PEO ultra-thin films with and without plasmon enhancement in the range 1045-1200 cm$^{-1}$. The pristine IR absorption spectrum of the 8-nm PEO film without plasmon enhancement is shown as the black line at the bottom. The vertical lines indicate the positions of modes E, F and G.

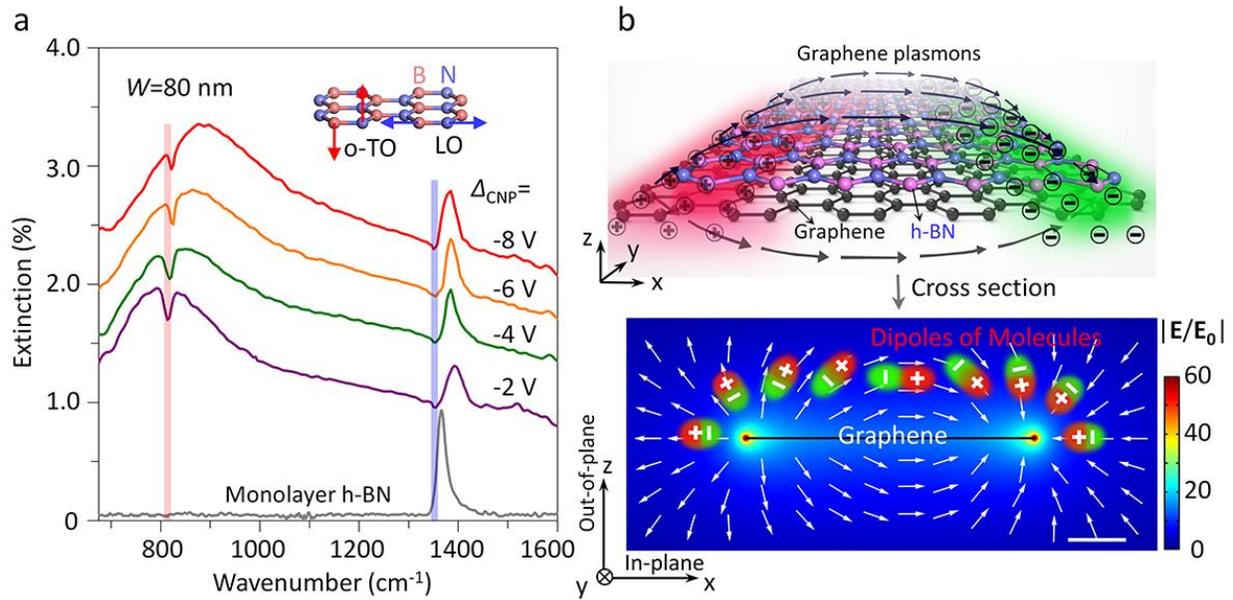

**Figure 5. Simultaneous detection of in-plane and out-of-plane vibrational fingerprints. (a)** The extinction spectra (colored lines) of the graphene plasmon sensor covered with an h-BN monolayer. The IR extinction spectrum (grey line) of monolayer BN is obtained with incident light normal to the h-BN basal plane. The vertical lines indicate the positions of the optical phonon modes of h-BN monolayer. Inset: the out-of-plane (the transverse optical phonon mode at approximately 820 cm⁻¹, o-TO) and in-plane (the longitudinal optical phonon mode at approximately 1370 cm⁻¹, LO) modes in h-BN. **(b)** A schematic diagram of the interaction between the electric field of the graphene plasmon and monolayer h-BN structure vibrations. The red and green colors in the upper figure indicate the snapshot of positive and negative charge distribution of graphene plasmon. The black arrows represent graphene plasmon. The lower part shows the side view of the electric field intensity distribution calculated from 100 nm wide graphene nanoribbons with $E_F = 0.3$ eV, obtained from a finite element electromagnetic simulation. White arrows indicate the relative direction of the distribution of electric field of graphene plasmon and the response of molecular vibrations to the plasmonic electric field is illustrated by dipoles. The color bar indicates the field confinement of graphene plasmon, while $\mathbf{E_0}$ is the electric field intensity of incident light. Scale bar, 20 nm.

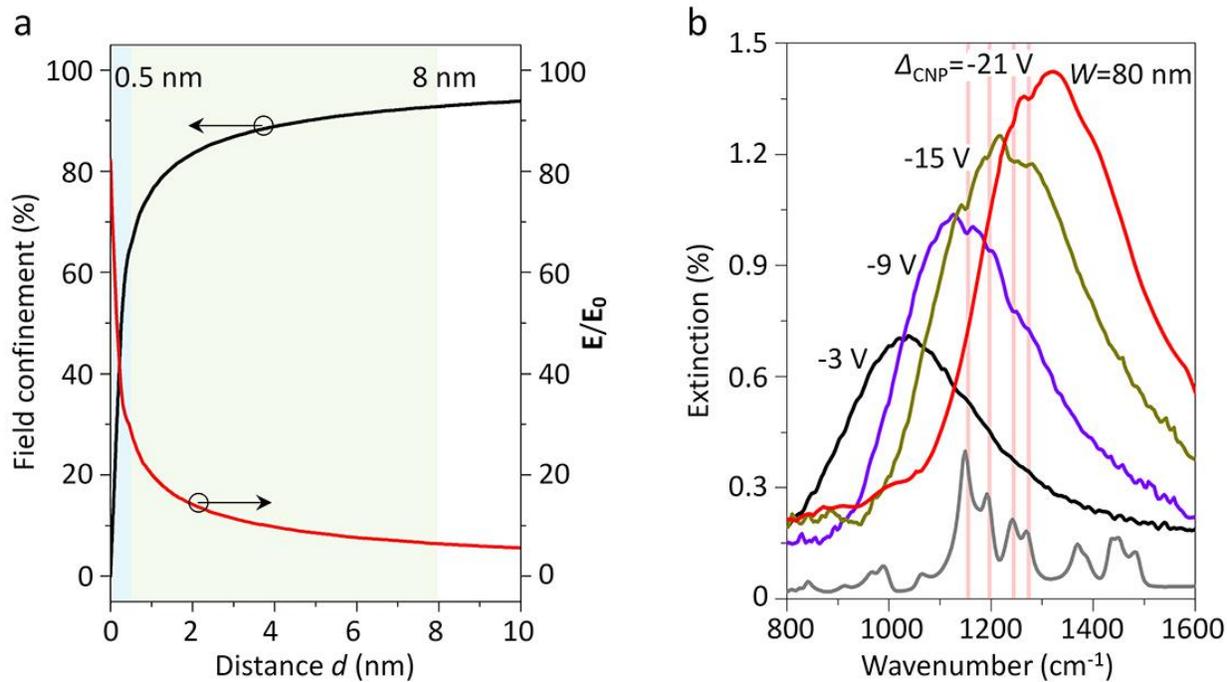

**Figure 6. Sub-monolayer polymer molecules detection.** (a) Near-field intensity confinement and near-field enhancement as a function of the distance $d$ from the graphene nanoribbons. The shaded regions correspond to $d$ less than 0.5 nm and 8 nm, respectively. $\mathbf{E_0}$ is the electric field intensity of incident light. (b) The sub-monolayer residual PMMA polymer extinction spectra for molecular vibrational fingerprinting at different effective gate voltages. An absorption spectrum of a 300-nm PMMA film is provided to indicate the PMMA vibrational fingerprint modes. Vertical lines indicate four strong vibrational fingerprints of PMMA.